\documentclass[review,times,authoryear]{elsarticle}
\usepackage{graphicx}
\usepackage{amssymb}
\usepackage{multirow}
\begin{document}

\title{Iron spin crossover and its influence on post-perovskite transitions in MgSiO$_3$ and MgGeO$_3$ }

\author[spa]{Gaurav Shukla}\corref{cor1}
\ead{shukla@physics.umn.edu}
\author[cems]{Mehmet Topsakal}
\ead{mtopsaka@umn.edu}
\author[spa,cems,msi,titj]{Renata M. Wentzcovitch}
\ead{wentz002@umn.edu}

\cortext[cor1]{Corresponding author}
\address[spa] {School of Physics and Astronomy, University of Minnesota, Minneapolis, Minnesota 55455, USA}

\address[cems]{Department of Chemical Engineering and Materials Science, \\
                       University of Minnesota, Minneapolis, Minnesota 55455, USA}

\address[msi]{Minnesota Supercomputer Institute, University of Minnesota, Minneapolis, Minnesota 55455, USA} 

\address[titj]{Earth-Life Science Institute, Tokyo Institute of Technology, 2-12-1 OoKayama, Meguro-ku, Tokyo, 152-8550, Japan}

\date{\today}

\begin{abstract}
MgGeO$_3$-perovskite is known to be a low-pressure analog of MgSiO$_3$-perovskite in many respects,
but especially in regard to the post-perovskite transition.
As such, investigation of spin state changes in Fe-bearing MgGeO$_3$ might help to clarify some 
aspects of this type of state change in Fe-bearing MgSiO$_3$. Using DFT+U calculations, we have 
investigated pressure induced spin state changes in Fe$^{2+}$ and Fe$^{3+}$ in MgGeO$_3$ perovskite and post-perovskite.
Owing to the relatively larger atomic size of germanium compared to silicon, germanate phases
 have larger unit cell volume and inter-atomic distances than equivalent silicate phases at 
same pressures. As a result, all pressure induced state changes in iron occur at higher 
pressures in germanate phases than in the silicate ones, be it a spin state change or 
position change of (ferrous) iron in the perovskite cage. We showed that iron state transitions occur at particular average Fe-O bond-length irrespective of mineral composition
(silicate or germanate) or functionals (LDA+U$_{sc}$ or GGA+U$_{sc}$). 
Ferrous iron substitution decreases the perovskite to post-perovskite (PPv) transition pressure while coupled ferric iron substitution increases it noticeably. 

\end{abstract}


\maketitle

\section{Introduction}

Iron-bearing MgSiO$_3$ perovskite (Si-Pv) is one of the major constituent minerals of Earth's lower mantle
along with (Mg,Fe)O ferropericlase (Fp). Unraveling the composition and thermal structure of the lower mantle
requires a detailed understanding of the influence of iron (Fe) on these host minerals. 
It is well known that iron in (Mg,Fe)O undergoes a pressure induced spin crossover from high-spin (S=2) to
low-spin (S=0) state and affects elastic properties \citep{Badro03,Badro04,Goncharov,Crowhurst,Tsuchiya06,Marquardt09a,Wu09,Antonangeli11,Wu13,Hsu14,Wu14}.
In the case of Si-Pv, deciphering the iron spin state has been quite challenging due to complex perovskite structure and
possibility of different valence states of iron (ferric and ferrous). 
Si-Pv is also known to undergo a perovskite (Pv) to post-perovskite (PPv) phase transition at $\sim$ 125 GPa  
and this will have geophysical consequences in deep lower mantle region \citep{Murakami04, Tsuchiya04, Oganov04}.
Understanding the effect of iron, especially, on Pv to PPv transition and on elastic properties of 
Pv and PPv phases is crucial to constrain lower mantle composition.

In (Mg,Fe$^{2+}$)SiO$_3$-Pv, the pressure induced increase in M$\ddot{o}$ssbauer quadrupole splitting (QS) above $\sim$30 GPa (low-QS to high-QS transition)
\citep{Jackson05, Li04, Li06,McCammon08, McCammon10, Lin12, Lin13} has been explained by lateral displacement of Fe$^{2+}$ ion which remains in the high-spin (HS) state in 
the entire lower-mantle pressure range \citep{Bengtson09, Hsu10, Hsu14}. In the case of (Mg,Fe$^{3+}$)(Si,Fe$^{3+}$)O$_3$-Pv, however, 
first-principles calculations by \citet{Hsu11} reported that Fe$^{3+}$ in B-site undergoes a high-spin (S=5/2) to low-spin (S=1/2) state change at approximately 41 GPa and 
70 GPa using LDA+U$_{sc}$ and  GGA+U$_{sc}$ methods, respectively, while in the A-site Fe$^{3+}$ remains in the high-spin state. These findings 
were in good agreement with experimental observations of \citet{Catalli10b}. There have also been several experimental \citep{Lin08,Catalli10a,Mao10} and
first-principles \citep{Zhang_and_Oganov06,Caracas08, Yu12}
studies about the spin crossover in  MgSiO$_3$-PPv (Si-PPv). The consensus regarding the spin crossover
in iron-bearing Si-PPv is that Fe$^{2+}$ in the A-site remains in the high-spin
and high-QS state throughout the lower mantle pressure range, while Fe$^{3+}$ undergoes a high to low spin transition in the B site and remains in the high spin state in the A-site \citep{Yu12}. In spite
of these efforts to pin down the state of iron at high pressures, it has been very difficult to ascertain the exact role of iron with its varying spin and valence state on Pv to PPv transition,
especially because of the extreme pressure and temperature conditions in the deep lower mantle region. 

MgGeO$_3$-perovskite (Ge-Pv) is a low pressure analog of MgSiO$_3$ perovskite \citep{Ross88,Leinenweber94,Hirose05,Kubo06,Merkel06,Runge06,Shim07,Tsuchiya07,Duffy08,Ito10}
and would be a promising candidate to 
investigate iron spin crossover and its structural influence on MgSiO$_3$-Pv and PPv phases. In this study, we have 
used first-principles DFT+U (LDA+U$_{sc}$ and GGA+U$_{sc}$) calculations to investigate the iron spin states and their crossover in the Pv 
and PPv phases of Fe$^{2+}$- and Fe$^{3+}$-bearing MgGeO$_3$. Comparing enthalpy of the different spin states, 
we have determined the stability of the different states and transition pressure. Using these results, we have also
investigated the effect of Fe$^{2+}$ and Fe$^{3+}$ substitution on Pv to PPv transition.

\section{Method}
We have used density functional theory (DFT) within the local density approximation (LDA) \citep{ceperley}  and generalized gradient 
approximation (GGA) \citep{Perdew96} augmented by the Hubbard U (LDA/GGA+U$_{sc}$), calculated 
self-consistently and structurally consistently \citep{Cococcioni,Kulik, Hsu09,Hsu11,Himmetoglu14}.
Ultrasoft pseudo-potentials, generated by Vanderbilt's method \citep{Vanderbilt},
have been used for Fe, Si, Ge, and O. For Mg, norm-conserving pseudo-potential generated by von Barth-Car's method, has been used. 
The plane-wave kinetic energy and charge density cutoff are 50 Ry and 300 Ry, respectively.
Self-consistent U$_{sc}$, calculated by \citet{Hsu10} and \citet{Yu12} using the linear response approach \citep{Cococcioni,Kulik} (Table \ref{table:U_sc}), have been used.
All calculations were performed in a 40-atoms super-cell. Electronic states were sampled on a shifted $2\times2\times2$ and
$3\times2\times3$ k-point grid for Pv and PPv structure, respectively \citep{Monkhorst_Pack}. Structural optimization at any arbitrary
volume has been performed using variable cell-shape damped molecular dynamics approach \citep{Wentzcovitch91,Wentzcovitch93}.
Structures are optimized until the inter-atomic forces are smaller than 10$^{-4}$ $Ry/a.u$.
For every possible spin and valence state, structures are optimized at twelve volumes in the relevant pressure-range 
and the 3$^{rd}$-order Birch-Murngharn equation of state has been used to fit these compression data.

\section{Iron spin states in Pv and PPv phases}
Iron spin states have been investigated in Fe$^{2+}$ and Fe$^{3+}$-bearing MgGeO$_3$ and MgSiO$_3$ in a 40-atom super-cell. One Fe$^{2+}$ ion is
substituted for a Mg$^{2+}$ ion in the A-site in (Mg$_{0.875}$Fe$_{0.125}$)GeO$_3$ and (Mg$_{0.875}$Fe$_{0.125}$)SiO$_3$ while two Fe$^{3+}$'s are substituted for one Mg$^{2+}$ and
one first neighbor Ge$^{4+}$ or Si$^{4+}$ as in (Mg$_{0.875}$Fe$_{0.125}$)(Ge$_{0.875}$Fe$_{0.125}$)O$_3$ and (Mg$_{0.875}$Fe$_{0.125}$)(Si$_{0.875}$Fe$_{0.125}$)O$_3$. 
The nearest-neighbor Fe$^{3+}$-Fe$^{3+}$ substitution has been chosen as it has been found to be the lowest energy configuration in previous first-principles studies \citep{Hsu11,Hsu12,Yu12}.
Relative enthalpies have been used to calculate 
transition pressures (P$_t$) and the stability of a state or phase in a given pressure range. Effects of disorder have not been addressed at this point,
but it is expected to produce a two-phase loop. Here we wish to analyze (a) the potentially different effects of Fe$^{2+}$ and Fe$^{3+}$ in the Pv to PPv transition and (b)
whether the germanate phase could serve as a good low pressure analog to investigate spin transitions in iron.

\subsection{Spin states of Fe$^{2+}$ in Ge-Pv}
\begin{figure}
\includegraphics[width=8cm]{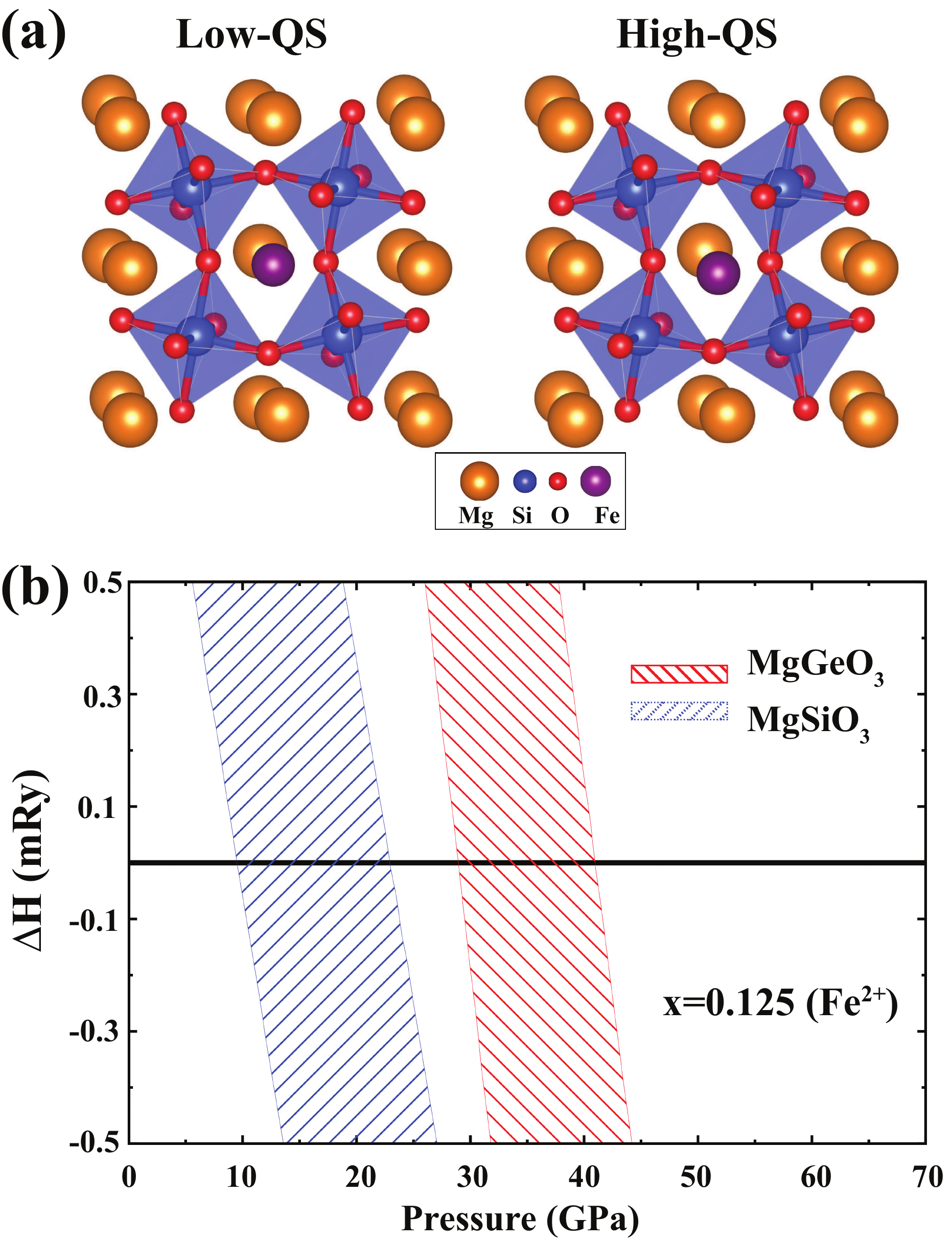}
\caption{(Color Online) (a) Atomic structure around low-QS and high-QS  Fe$^{2+}$-bearing MgSiO$_3$ in Pv phase at ambient pressure. Fe, Mg, O, and Si are represented as purple, orange, red, and blue sphere, respectively. MgGeO$_3$-Pv structures are similar and not shown here. (b) Enthalpy difference ($\Delta$H=H$_{high-QS}$-H$_{low-QS}$) for 
Fe$^{2+}$-bearing Si-Pv (blue) and  Ge-Pv (red).
Lower and upper bounds correspond to LDA+U$_{sc}$ and GGA+U$_{sc}$ results, respectively. }
\label{fig1}
\end{figure}

As in the case of Fe$^{2+}$in Si-Pv \citep{Bengtson09,Hsu10, Hsu14}, Fe$^{2+}$ in Ge-Pv 
remains in the high-spin (HS, S=2) state in the entire lower-mantle pressure range (0-135 GPa). However, it undergoes a pressure induced lateral displacement producing a transition from low to high M$\ddot{o}$ssbauer quadrupole splitting (QS). The atomic structures for the low- and high-QS states of iron in Si-Pv are shown in Fig. \ref{fig1}(a) indicating the lateral displacement of iron in the perovskite A-site. 

The pressure dependence of enthalpy differences between high-QS and low-QS states ($\Delta$H=H$_{high-QS}$-H$_{low-QS}$) gives the low- to high-QS transition pressure. It is well-known that LDA usually underestimates transition pressures while GGA usually overestimates it. In this study, we present both LDA+U$_{sc}$ and GGA+U$_{sc}$ results to provide lower and upper bounds for this transition, as shown in Fig. \ref{fig1}(b). The low- to high-QS state change in Ge-Pv occurs at $\sim$28.5 GPa with LDA+U$_{sc}$ and at $\sim$40.5 GPa with GGA+U$_{sc}$. These transition pressures are quite large compared to those in Si-Pv, i.e., $\sim$9.5 GPa and $\sim$22.5 GPa respectively as presented in Table \ref{table:P_t}.

\begin{figure}
\includegraphics[width=8cm]{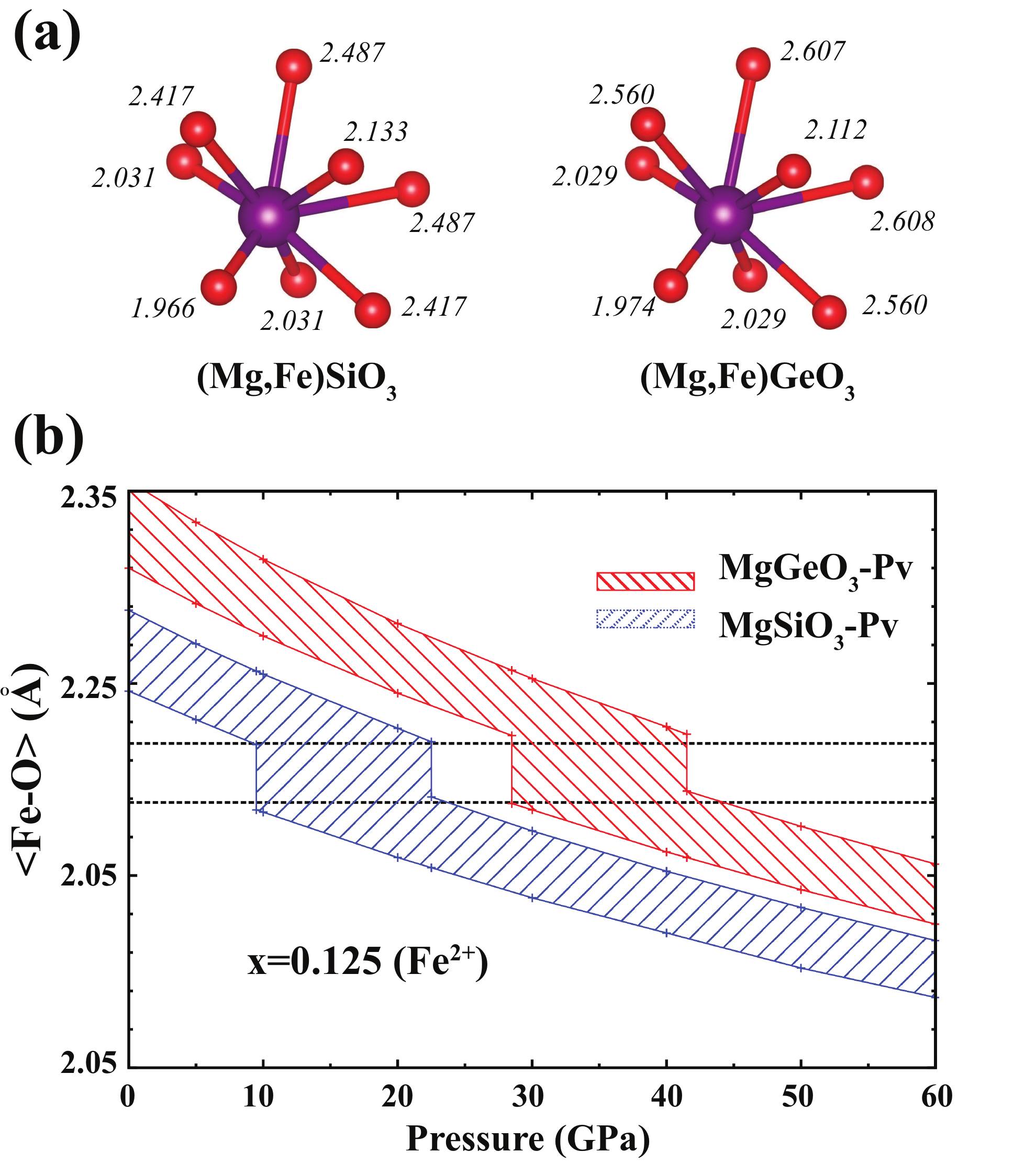}
\caption{(Color Online)  Atomic environment around iron in $Fe^{2+}$-bearing Si-Pv and Ge-Pv phases. (a) Fe-O distances (in $\mathring{A}$) in  Si-Pv, and  Ge-Pv structures at
ambient pressure are shown next to the oxygen atoms. (b)  Pressure dependence of average Fe-O distance. Vertical lines represent the low-QS to high-QS transition. Lower and upper bounds correspond to LDA+U$_{sc}$ and GGA+U$_{sc}$ results, respectively.}
\label{fig2}
\end{figure}

Owing to the larger ionic radius of Ge$^{4+}$ (0.53 $\mathring{A}$) compared to that of Si$^{4+}$ (0.40 $\mathring{A}$), the unit-cell volume and inter-atomic distances in Ge-Pv are larger than in Si-Pv. Zero pressure Fe-O bond-lengths in Si-Pv and Ge-Pv obtained with LDA+U$_{sc}$ are shown Fig. \ref{fig2}(a). 
The average of Fe-O bond-lengths, $\langle$Fe-O$\rangle$, is also shown in Fig. \ref{fig2}(b). The larger $\langle$Fe-O$\rangle$ value in Ge-Pv allows Fe$^{2+}$ to sustain higher pressures than in Si-Pv before changing to the high-QS state. It is also interesting to note that this transition occurs when the $\langle$Fe-O$\rangle$ reaches a particular value ($\sim$2.22 $\mathring{A}$) whether in Ge-Pv or in Si-Pv, whether using LDA+U$_{sc}$ or GGA+U$_{sc}$ methods (Fig. \ref{fig2}.(b)). It should be noted that the Pv to PPv transition is expected to occur at approximately $\sim$63 GPa in MgGeO$_3$ \citep{Hirose05} and at $\sim$125 GPa in MgSiO$_3$ \citep{Murakami04}. These pressures are 
much higher than low- to high-QS transition pressures in both Fe$^{2+}$-bearing compounds. Consequently, this transition has not been investigated in Si-PPv or Ge-PPv phases.

\subsection{Spin states of Fe$^{3+}$ in Ge-Pv and -PPv}
\begin{figure}
\includegraphics[width=8cm]{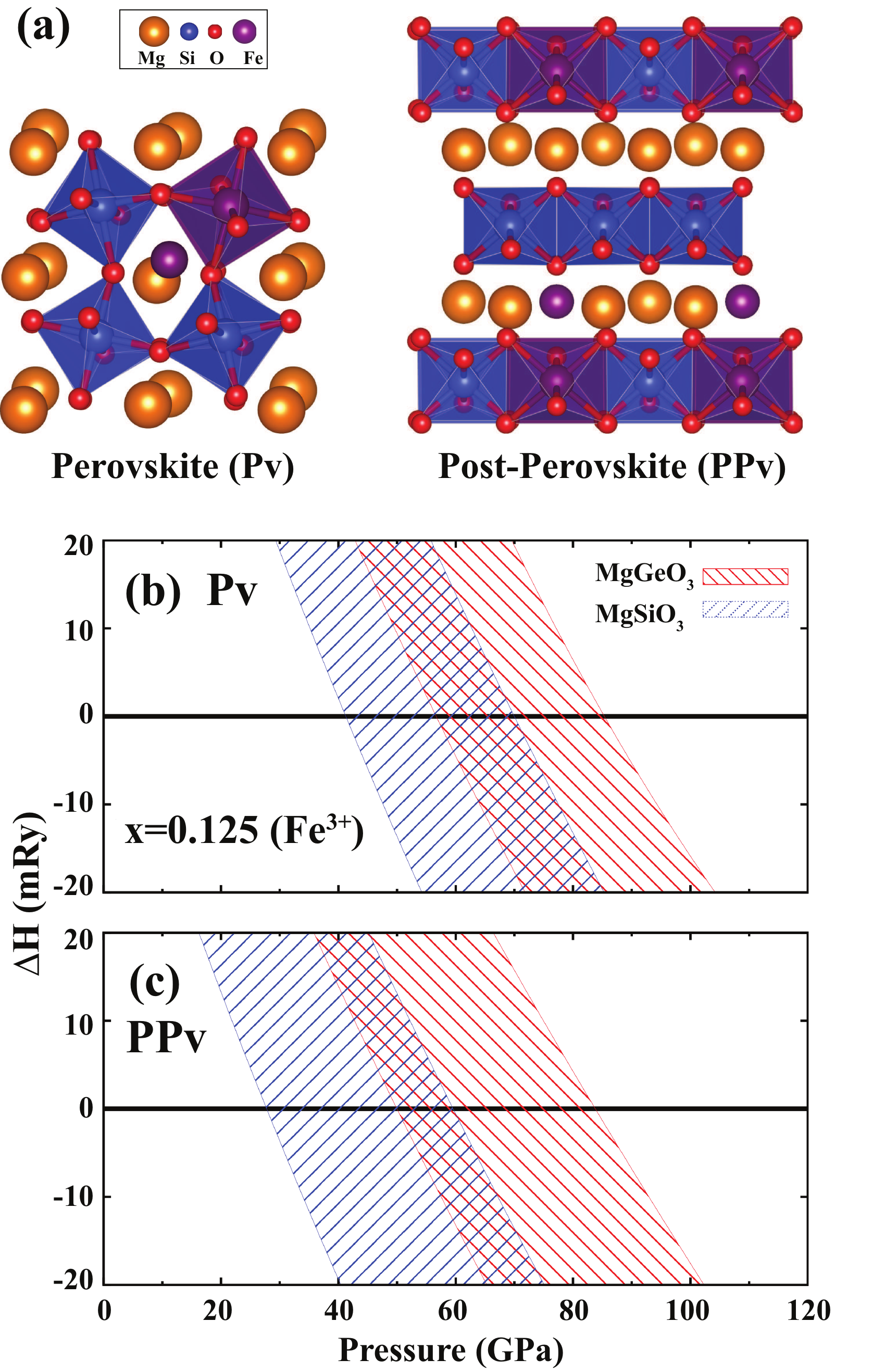}
\caption{(Color Online) (a) Atomic structure of Fe$^{3+}$-bearing MgSiO$_3$ Pv and PPv phase. Structures of Pv and PPv phases of Fe$^{3+}$-bearing MgGeO$_3$  are similar and not shown here.
Fe, Mg, O, and Si are represented as purple, orange, red, and blue sphere, respectively.  Enthalpy difference ($\Delta$H=H$_{B(LS)}$-H$_{B(HS)}$)  
for Fe$^{3+}$-bearing MgSiO$_3$ (blue) and MgGeO$_3$ (red) in (b) Pv and (c) PPv phases. Lower and upper bounds correspond to LDA+U$_{sc}$ and GGA+U$_{sc}$ results, respectively.}
\label{fig3}
\end{figure}

Next, we investigate spin states of Fe$^{3+}$ in the B-site of Ge-Pv (low-pressure) and PPv (high-pressure) phases. 
The crystal structures of Si-Pv and Si-PPv phases are shown in Fig. \ref{fig3}(a). 
Si-PPv has a layered structure characterized by corner-sharing and edge-sharing SiO$_6$ octahedra extending along the a-c plane. 
The pressure dependence of calculated enthalpy differences between high-spin (HS) and low-spin (LS) states of Fe$^{3+}$ ($\Delta$H=H$_{B(LS)}$-H$_{B(HS)}$) in Pv and PPv phases are shown in Fig. \ref{fig3} (b) and (c), respectively.
The calculated HS to LS transition pressures in Ge-Pv are 56.5 GPa with LDA+U$_{sc}$ and 85.0 GPa with GGA+U$_{sc}$. Similarly, these transition pressures in Ge-PPv are 49.5 GPa and 82.5 GPa with these methods.
Our calculated transition pressures in Si-Pv are 41.0 GPa with LDA+U$_{sc}$ and 69.5 GPa with GGA+U$_{sc}$, which are in very good agreement with previous first-principles study \citep{Hsu11}. 
In the case of Si-PPv, our calculated transition pressures are 28.0 GPa and 59.5 GPa, respectively, which are quite larger than previously reported values of -30 GPa and 36 GPa \citep{Yu12}. 
These transition pressures in Si-Pv and Si-PPv are also very similar to those reported by \citet{Hsu12} in (Mg,Al)(Si,Fe)O$_3$-Pv and PPv phases where it was observed that the presence of aluminum does not affect the spin transition pressure significantly. 
Considering the overall agreement of the stability of the iron spin states and  pressure induced crossover in Si-Pv and Si-PPv phases with previous studies \citep{Hsu10, Hsu11, Hsu12}, the present calculations about HS to LS transition in Fe$^{3+}$ at 
B site seems to be consistent and reliable.

\begin{figure}
\includegraphics[width=8cm]{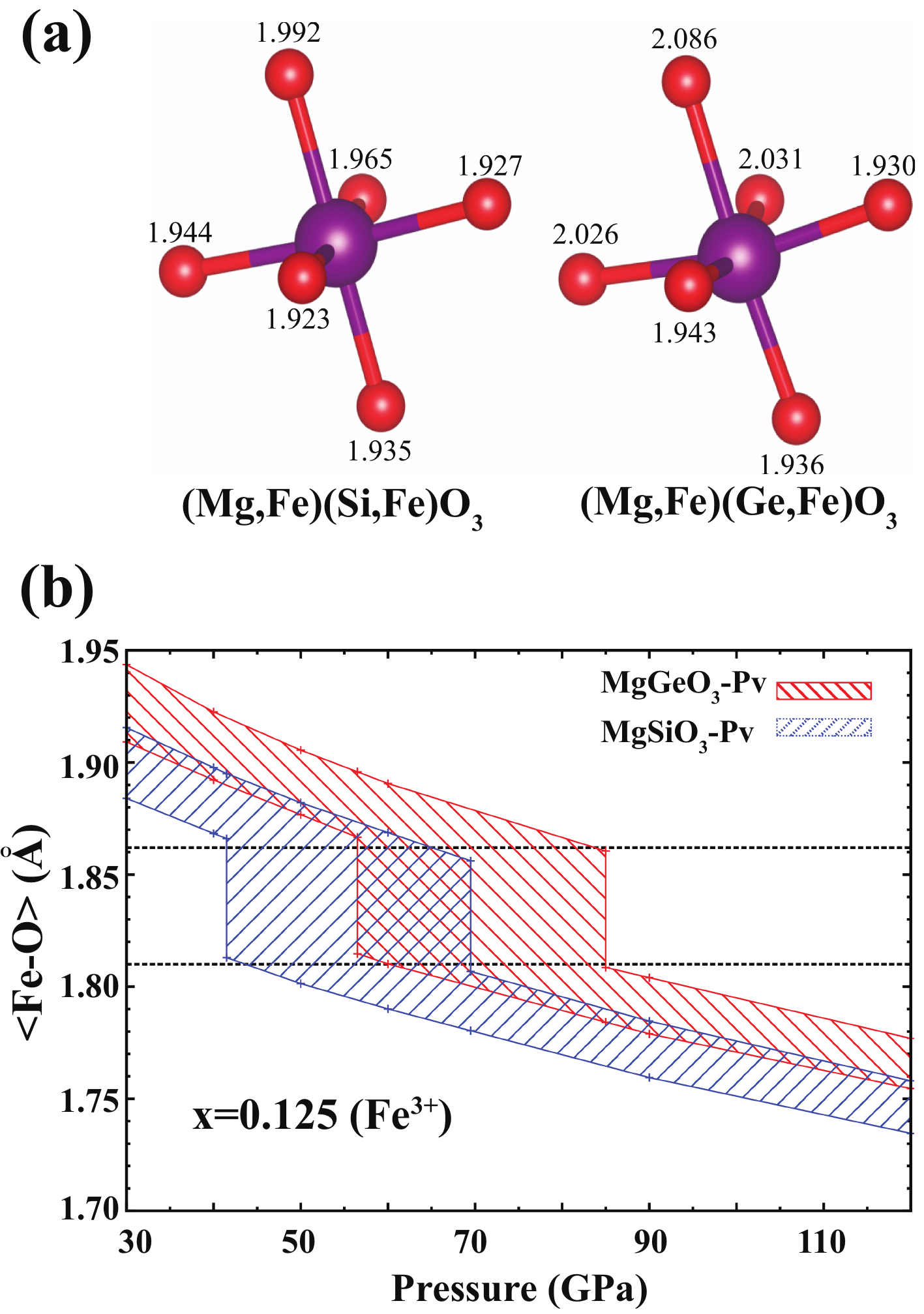}
\caption{(Color Online) Atomic environment around $Fe^{3+}$ in Si-Pv and Ge-Pv phases. (a)  Zero pressure Fe-O bond lengths ($\mathring{A}$) in  Si-Pv and in Ge-Pv obtained with LDA+U$_{sc}$. (b) Pressure dependence of average Fe-O bond lengths for MgSiO$_3$ (blue) and MgGeO$_3$ (red) phases. Vertical lines represent the high- (HS) to low-spin (LS) transitions in the B-site. Lower and upper bounds correspond to LDA+U$_{sc}$ and GGA+U$_{sc}$ results, respectively.}
\label{fig4}
\end{figure}

In order to explain the increase of HS to LS transition pressures in MgGeO$_3$ phases, we compare atomic configurations around Fe$^{3+}$ in the B-site of Si-Pv and Ge-Pv phases.  
As shown in Fig. \ref{fig4}(a), the coordination number of Fe$^{3+}$ is six and Fe-O bond lengths in this site are smaller than those of Fe$^{2+}$ in the A-site. 
The pressure dependence of the average bond-length, $\langle$Fe-O$\rangle$, for Fe$^{3+}$ in the B-site in Si-Pv and Ge-Pv phases are shown in Fig. \ref{fig4}(b). 
As in the case of low- to high-QS transition in Fe$^{2+}$
(Fig. \ref{fig1}), the HS to LS transition in Fe$^{3+}$ in Ge-Pv 
and Ge-PPv phases are also large compared to those in Si-Pv and Si-PPv phases (Fig. \ref{fig3}) due to relatively larger Fe-O bond-lengths and unit-cell volumes in germanate phases at a given pressure (Fig. \ref{fig4}). Similar to low- to high-QS transition, the HS to LS transition also occurs only when the $\langle$Fe-O$\rangle$ reaches a particular value (i.e., $\sim$1.86 $\mathring{A}$ in the present case)
irrespective of Ge-Pv or Si-Pv, or of the exchange-correlation functional used (Fig. \ref{fig4}(b)). This is a useful insight for predicting pressure induced spin state changes in different materials.

\section{Effect of Fe$^{2+}$ and Fe$^{3+}$ in the Pv to PPv transition}

% we need to ad some story here 
%
Having investigated the stability of different iron states in the lower mantle pressure range in the previous section, we now focus on the potential effect of Fe$^{2+}$ and Fe$^{3+}$ substitution on the Pv to PPv transition. This is a plain static calculation to access the overall influence of iron substitution on the Pv to PPv transition in these phases. We have not included finite temperature entropic or vibrational effects, therefore, we are not able to produce a full phase diagram with a two phase loop region. This is simply the first step towards the calculation of such phase diagrams. 

It is well known that the LDA functional underestimates the equilibrium volume while GGA overestimates it. After inclusion of vibrational effects LDA equations of state and elastic properties tend to agree better with experimental data than GGA results \citep{Karki99, Karki00,Wentzcovitch04,Nunez-Valdez12a, Nunez-Valdez12b,Nunez-Valdez13}. Therefore, we have chosen LDA+U$_{sc}$ results to investigate the effect of Fe$^{2+}$ and Fe$^{3+}$ in the Pv to PPv transition in both compounds. Previous studies based on 
\textit{ab initio} calculations \citep{Tsuchiya04, Tsuchiya07} and experimental measurements \citep{Murakami04,Hirose05} have reported the Pv to PPv transition in MgSiO$_3$ and MgGeO$_3$ to occur around $\sim$125 GPa and $\sim$63 GPa, respectively, at $\sim$2500 K.

\begin{figure}
\includegraphics[width=8cm]{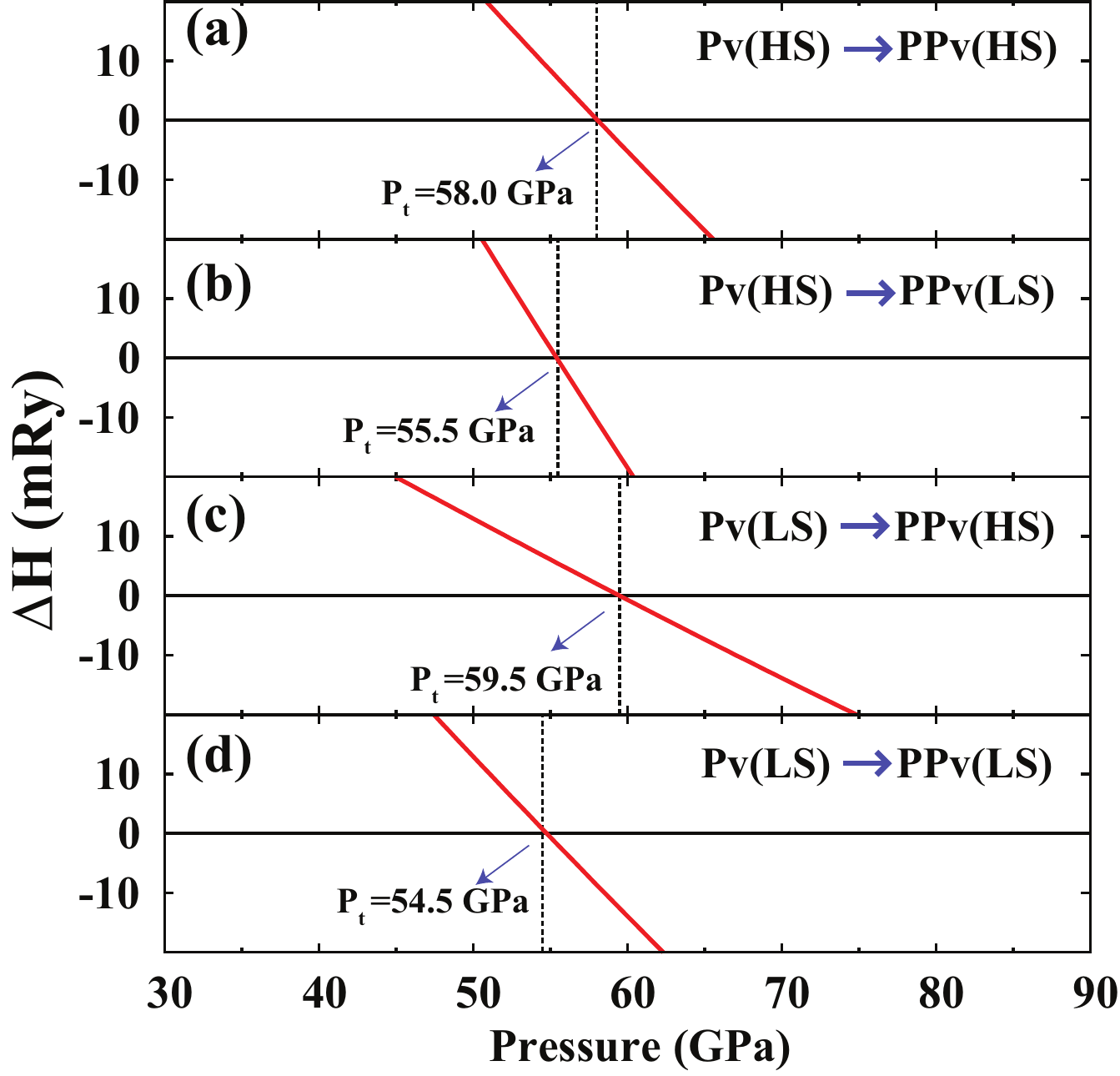}
\caption{(Color Online) Enthalpy difference ($\Delta$H=H$_{PPv}$-H$_{Pv}$) between Fe$^{3+}$-bearing Ge-Pv and Ge-PPv phase using LDA+U$_{sc}$ method with different possible spin-state combinations
of Fe$^{3+}$ at B-site. (a) Pv(HS) and PPv(HS), (b) Pv(HS) and PPv(LS), (c) Pv(LS) and PPv(HS), (d) Pv(LS) and PPv(LS).}
\label{fig5}
\end{figure}

As indicated in Fig. \ref{fig1}(b) and in Table \ref{table:P_t}, the low- to high-QS transition in Si-Pv and Ge-Pv occurs at much lower pressures than the Pv to PPv transition in these minerals \citep{Murakami04,Hirose05}. 
Considering this fact, only the high pressure high-QS state of Fe$^{2+}$ was considered in the investigation of the Pv to PPv transition. 
Similarly, near the Pv to PPv transition pressure, the stable state of Fe$^{3+}$ substituted in both Si-Pv and Si-PPv phases is HS state in the A-site (HS-A) and LS state in the B-site (LS-B).
However, in germanate phases, the HS-B to LS-B transition pressure 
(Fig. \ref{fig3} and Table \ref{table:P_t}) is near the
Pv to PPv transition pressure. In order to find the appropriate spin states of Fe$^{3+}$ across the transition we have considered all possibilities in the calculation of enthalpy differences  ($\Delta$H=H$_{PPv}$-H$_{Pv}$) (see Fig. \ref{fig5}). As indicated in Fig. \ref{fig5}(b) the transition pressure 55.5 GPa involves a transition between HS-B in Ge-Pv to LS-B in Ge-PPv. Obviously iron in the A site remains in the HS state. At high temperatures, contributions by other states might be relevant as well, but this issue is well beyond the scope of this paper.
This transition and transition pressure, i.e., HS-B in Pv and LS-B in PPv, is the only combination consistent with B(HS) to B(LS) transition pressures of 56.5 GPa and 49.5 GPa in Ge-Pv and Ge-PPv, respectively.

\begin{figure}
\includegraphics[width=8cm]{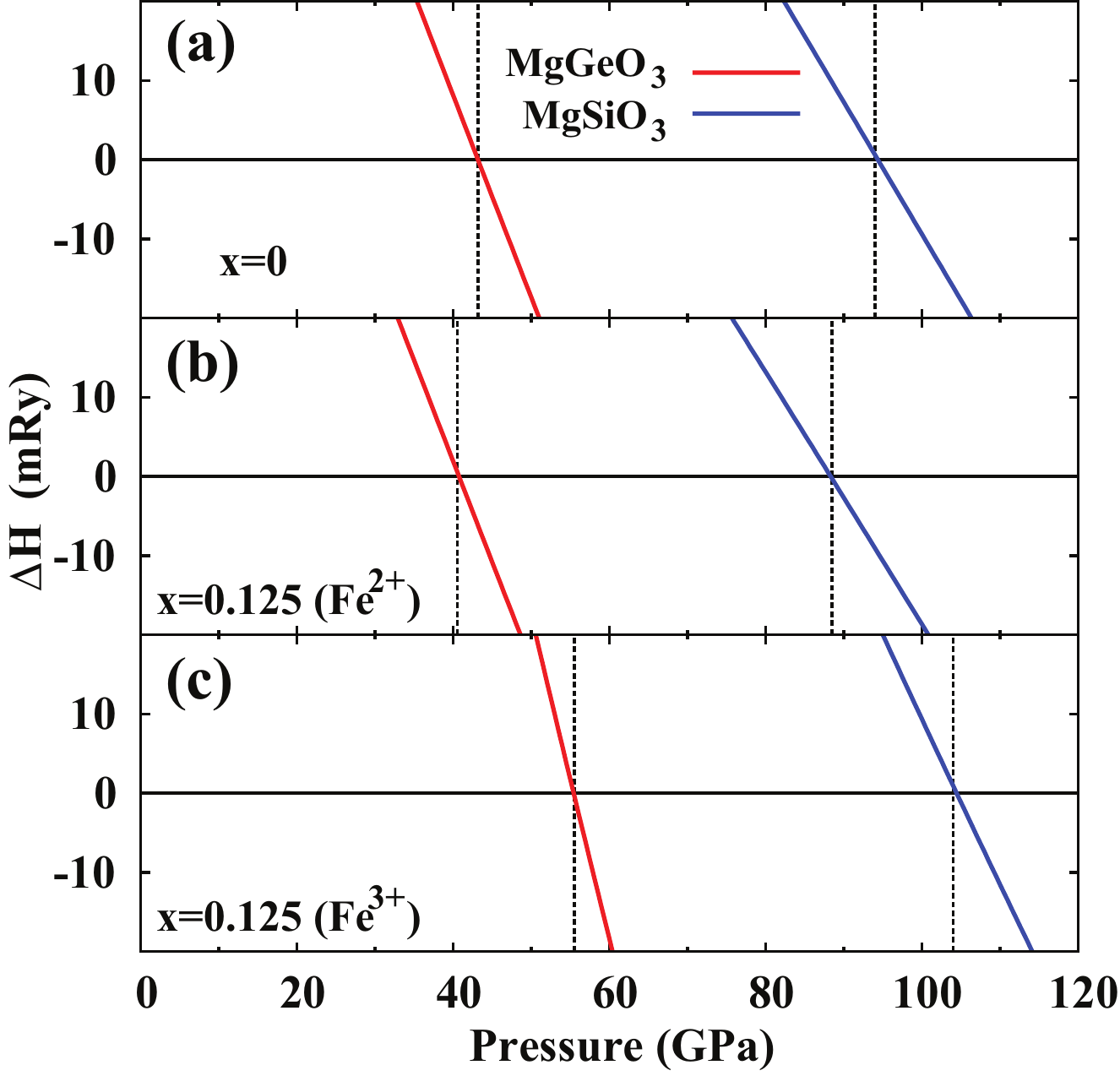}
\caption{(Color Online) Enthalpy difference ($\Delta$H=H$_{PPv}$-H$_{Pv}$) obtained using LDA+U$_{sc}$ for (a) iron free, (b) Fe$^{2+}$-
, and (c) Fe$^{3+}$-bearing Si-Pv and Ge-Pv.}
\label{fig6}
\end{figure}

The effect of Fe$^{2+}$ and Fe$^{3+}$ on the Pv to PPv transition in Si-Pv and Ge-Pv is summarized in Fig. \ref{fig6} showing the pressure dependence of enthalpy differences ($\Delta$H=H$_{PPv}$-H$_{Pv}$) for the configurations indicated in the previous paragraph. Static (zero temperature) Pv to PPv transition pressures in iron-free (x=0) MgGeO$_3$ and MgSiO$_3$ are 43.2 GPa and 94 GPa, respectively. The Clapeyron slope for the Pv to PPv transitions have been previously estimated to be $\sim$ 7.5 MPa/K for MgSiO$_3$ \citep{Tsuchiya04} and 7.8 MPa/K for MgGeO$_3$ \citep{Tsuchiya07}. 
Taking these Clapeyron slopes approximately into account, transition pressures at $\sim$2500 K should be $\sim$ 62.7 GPa and $\sim$ 112.8 GPa for MgGeO$_3$ and MgSiO$_3$, respectively. These results are in good agreement with previous experimental measurements and \textit{ab initio} LDA calculations \citep{Murakami04, Hirose05, Tsuchiya04, 
Tsuchiya07}.

A summary of transition pressure shifts caused by iron in silicate and germanate phases is indicated in (Fig. \ref{fig6}, 
Table \ref{table:Pv-PPv-P_c}).  As indicated, Fe$^{2+}$ decreases the Pv to PPv transition pressure in both MgGeO$_3$ and MgSiO$_3$. These observations are consistent with previous experimental \citep{Mao04,Shieh06} and \textit{ab initio} studies \citep{Caracas05, Stackhouse06}. However, in the case
Fe$^{3+}$, the Pv to PPv transition pressure increases noticeably. To the best of our knowledge, this fact has not been verified experimentally yet.

\begin{figure}
\includegraphics[width=8cm]{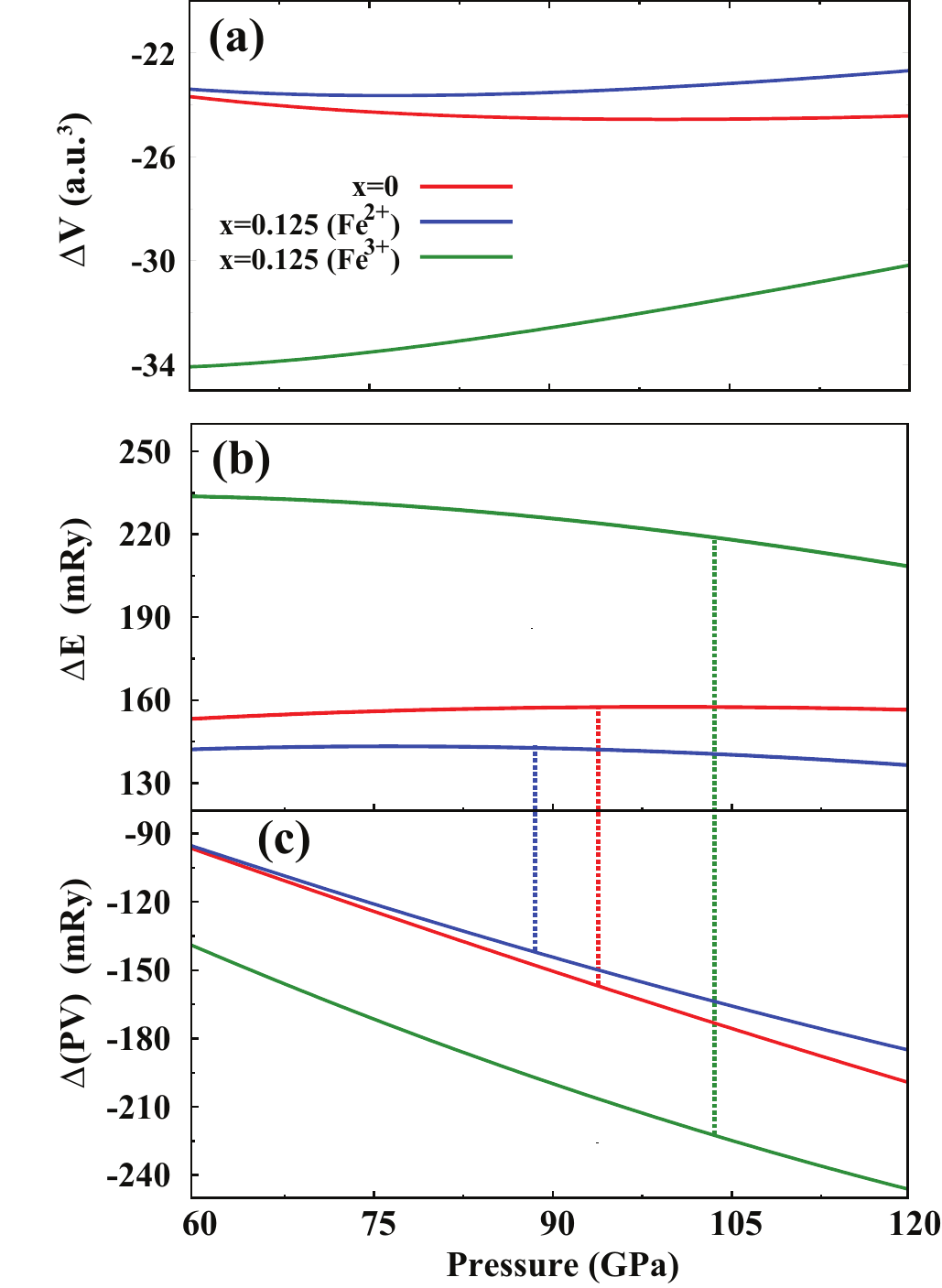}
\caption{(Color Online) Difference in (a) volume ($\Delta$V=V$_{PPv}$-V$_{Pv}$), (b) internal energy ($\Delta$E=E$_{PPv}$-E$_{Pv}$), and (c) PV term ($\Delta$(PV)=(PV)$_{PPv}$-(PV)$_{Pv}$) of the enthalpy for iron-free (red),  Fe$^{2+}$-bearing (blue), and  Fe$^{3+}$-bearing (green) MgSiO$_3$. Results represent LDA+U$_{sc}$ calculation. Vertical dashed lines represent Pv to PPv transition point.}
\label{fig7}
\end{figure}

To understand these opposite effects of Fe$^{2+}$ and Fe$^{3+}$ substitution on the Pv to PPv transition, the volume difference between Si-Pv and Si-PPv phase ($\Delta$V=V$_{PPv}$-V$_{Pv}$) is shown in Fig. \ref{fig7}(a). With respect to iron-free case, the volume difference  is smaller and larger in magnitude with Fe$^{2+}$ and Fe$^{3+}$ substitution, respectively. These smaller and larger volume difference will contribute to enthalpy difference between Pv and PPv phase in such a way that will decrease the trasniton pressure with Fe$^{2+}$  substitution while it will increase when Fe$^{3+}$ is substituted.
This can be seen explicitly in Fig. \ref{fig7}(b and c), where pressure dependence of energy ($\Delta$E) and P$\times$V term  [$\Delta$(PV)] of relative enthalpy ($\Delta$H) are shown separately. Transition pressure points, where $\Delta$E and $\Delta$(PV) terms cancel each other, are shown by vertical dashed lines.

\section{Conclusions}

Using \textit{ab initio} DFT+U calculations, we have undertaken a comparative investigation of state changes of ferrous (Fe$^{2+}$) and ferric (Fe$^{3+}$) iron in MgSiO$_3$ and MgGeO$_3$ perovskite (Pv) and post-perovskite (PPv) phases. The goal was to verify the possibility of using MgGeO$_3$ as a low pressure analog of MgSiO$_3$ for better understanding of spin state changes in iron. Unfortunately, owing to the larger ionic radius of germanium, Fe-O inter-atomic distances in the germanate phases are larger than in the silicate phases, allowing the germanates to sustain higher pressures than analog silicates before state changes in iron take place. It has been clearly shown that such state changes, whether the displacement transition (low- to high-QS) of Fe$^{2+}$ in the A-site or the HS to LS state change in Fe$^{3+}$ in the B-site, take place at approximately the same average Fe-O, $\langle$Fe-O$\rangle$, distance irrespective of the chemistry or the exchange and correlation functional used in the calculations.  

This encompassing investigation also allowed us to address the effect of 
Fe$^{2+}$ and Fe$^{3+}$ substitutions on the Pv to PPv transition. Despite the simplicity of these static calculations, previously measured and calculated effect of Fe$^{2+}$ on this transition in MgSiO$_3$ is well reproduced: Fe$^{2+}$ substitution decreases the transition pressure, whether in germanate or silicate phases. However, we report the opposite effect for Fe$^{3+}$ substitution in the B-site. This transition should involve the LS-B state in both Pv and PPv silicate phases but HS-B and LS-B states in Pv and PPv germanates, respectively. For both types of transitions the effect is the same: coupled substitution of Fe$^{3+}$ increases the Pv to PPv transition in both materials. This observation has not been validated by experiments yet and should be relevant for better understanding the D`` discontinuity and the nature of the deep lower mantle.

\section{Acknowledgments}
This work was supported primarily by grants NSF/EAR 1319368 and NSF/CAREER 1151738. 
Computations were performed at the Minnesota Supercomputing Institute (MSI) and at the Blue Waters System at NCSA. We thank Han Hsu
for helpful discussions.

\newpage

\begin{table}
\caption{Self-consistent Hubbard U$_{sc}$(eV) for ferrous and ferric iron calculated by \citet{Hsu10, Hsu11} in Pv and by \citet{Yu12} in PPv structures.}
\centering
\label{table:U_sc}
\begin{tabular}{ c | c c | c c c}
\hline
& \multicolumn{2}{c}{Fe$^{2+}$} &  \multicolumn{3}{c}{Fe$^{3+}$} \\ 
\hline
 & High-QS & Low-QS & A(HS) & B(HS) & B(LS) \\
Pv & 3.1& 3.1 & 3.7 & 3.3 & 4.9\\
PPv & 2.9 & -- & 4.0 & 3.5 & 5.6\\
\hline
\end{tabular}
\end{table}

\begin{table*}
\caption{Spin state transition pressures, P$_t$(GPa), in iron-bearing MgGeO$_3$ and MgSiO$_3$ calculated within LDA+U$_{sc}$ and GGA+U$_{sc}$ methods.} 
\centering
\label{table:P_t}
\begin{tabular}{ c | c c | c c | c c }
\hline
 &  \multicolumn{2}{|c|}{Low-QS to High-QS in Fe$^{2+}$-bearing Pv phase} &  \multicolumn{4}{|c}{B(HS) to B(LS) in Fe$^{3+}$-bearing Pv and PPv phases} \\
\hline
 &   \multicolumn{2}{|c|}{} & \multicolumn{2}{|c}{Pv} & \multicolumn{2}{|c}{PPv} \\
\hline 
 & LDA+U$_{sc}$ & GGA+U$_{sc}$ & LDA+U$_{sc}$ & GGA+U$_{sc}$ & LDA+U$_{sc}$ & GGA+U$_{sc}$ \\
MgGeO$_3$ & 28.5 & 40.5  & 56.5 & 85.0 & 49.5 & 82.5 \\
MgSiO$_3$ & 9.5 & 22.5  & 41.0 & 69.5 & 28.0 & 59.5\\
\hline
\end{tabular}
\end{table*}

\begin{table}
\caption{Perovskite to post-perovskite transition pressures, P$_C$(GPa), calculated within LDA+U$_{sc}$ methods. x represents the iron concentration.} \centering
\label{table:Pv-PPv-P_c}
\begin{tabular}{ c | c  c  c }
\hline
x & 0  & 0.125 (Fe$^{2+}$) & 0.125 (Fe$^{3+}$) \\
\hline
MgGeO$_3$ & 43.2  & 40.5  & 57.5 \\
MgSiO$_3$ & 94.0  & 88.5  & 104.0\\
\hline
\end{tabular}
\end{table}



\begin{thebibliography}{15}

\bibitem[{\textit{Antonangeli et~al.}(2011)\textit{Antonangeli, Siebert, Aracne, Farber, Bosak, Hoesch, Krisch, Ryerson, Fiquet, and Badro}}]
{Antonangeli11}
Antonangeli, D., J. Siebert, C.~M. Aracne, D.~L. Farber, A. Bosak, M. Hoesch, M. Krisch, F.~J. Ryerson, G. Fiquet, and J. Badro (2011), 
Spin Crossover in Ferropericlase at High Pressure: A Seismologically Transparent Transition?, \textit{Science}, \textit{331}, 64-67.

\bibitem[{\textit{Badro et~al.}(2003)\textit{Badro, Fiquet, Guyot, Rueff, Struzhkin, Vankó, Monaco}}]{Badro03}  Badro J., G. Fiquet, F. Guyot, J.~P. Rueff, V.~V. Struzhkin, G. Vankó, and G. Monaco (2003),
Iron Partitioning in Earth's Mantle: Toward a Deep Lower Mantle Discontinuity, \textit{Science}, \textit{300}, 789-791.


\bibitem[{\textit{Badro et~al.}(2004)\textit{Badro, Rueff, Vanko, Monaco, Fiquet, Guyot }}]
{Badro04}
Badro J., J.~P. Rueff, G. Vanko, G. Monaco, G. Fiquet, and F. Guyot (2004),
Electronic Transitions in Perovskite: Possible Nonconvecting Layers in the Lower Mantle, \textit{Science}, \textit{305}, 383.


\bibitem[{\textit{Bengtson et~al.}(2009)\textit{Bengtson, Li, and Morgan}}]{Bengtson09}
Bengtson, A., J.~Li, and D.~Morgan (2009), M$\ddot{o}$ssbauer modeling to interpret the spin sate of iron in (Mg,Fe)SiO$_3$, \textit{Geophys. Res. Lett.}, \textit{36}, L15301.

\bibitem[{\textit{Caracas et~al.}(2005)\textit{Caracas, and Cohen
  }}]{Caracas05}
 Caracas, R., and Cohen, R.~E., (2005), . Effect of chemistry on the stability and elasticity of the perovskite and post-perovskite phases in the MgSiO$_3$–FeSiO$_3$–Al$_2$O$_3$
system and implications for the lowermost mantle., \textit{Geophys. Res. Lett.}, \textit{32}, L16310.

\bibitem[{\textit{Caracas and Cohen}(2008)\textit{Caracas, and Cohen
  }}]{Caracas08}
 Caracas, R., and R.~E. Cohen (2008), Ferrous iron in post-perovskite from first-principles calculations, \textit{Phys. Earth Planet. Inter.}, \textit{168}, 147--152.
 
\bibitem[{\textit{Catalli et~al.}(2010a)\textit{Catalli, K., Shim, S.-H., Prakapenka, V.B., Zhao, J., Sturhahn, W., Chow, P., Xiao, Y., Liu, H., Cynn, H., Evans, W.J.
  }}]{Catalli10b}
 Catalli, K., S.-H. Shim, V.~B. Prakapenka, J. Zhao, W. Sturhahn, P Chow, Y. Xiao, H. Liu, H. Cynn, W.J. Evans (2010a), Spin state of ferric iron in MgSiO$_3$ perovskite and its
effect on elastic properties,  \textit{Earth Planet. Sci. Lett. }, \textit{289}, 68--75.
 
\bibitem[{\textit{Catalli et~al.}(2010b)\textit{Catalli, K., Shim, S.H., Prakapenka, V.B., Zhao, J., Sturhahn, W.
  }}]{Catalli10a}
  Catalli, K., S.-H. Shim, V.~B. Prakapenka, J. Zhao, W. Sturhahn (2010b), X-ray diffraction and Moessbauer spectroscopy of Fe$^{3+}$-bearing Mg-silicate post-perovskite at 128--138 GPa,
 \textit{American Mineralogist}, \textit{95}, 418--421.

 
\bibitem[{\textit{Ceperley and Alder}(1980)\textit{ceperley and Alder
  }}]{ceperley} 
 Ceperley, D. M., and B.~J. Alder (1980),  Ground state of the electron gas by a stochastic method, \textit{Phys. Rev. Lett.}, \textit{45}, 566--569. 
 
\bibitem[{\textit{Cococcioni and de Gironcoli}(2005)\textit{Cococcioni and de Gironcoli}}]{Cococcioni}
 Cococcioni, M., and S.~de Gironcoli (2005), Linear response approach to the calculation of the
effective interaction parameters in the LDA+U method, \textit{Phys. Rev. B}, \textit{71}, 035105.


\bibitem[{\textit{Crowhurst et~al.}(2008)\textit{Crowhurst, Brown, Goncharov, and Jacobsen
  }}]{Crowhurst}
 Crowhurst, J.~C., J.~M.~Brown, A.~F.~Goncharov, and S.~D.~Jacobsen (2008), Elasticity of (Mg,Fe)O through the spin transition of iron in the lower
mantle, \textit{Science}, \textit{319}, 451--453.

\bibitem[{\textit{Duffy}(2008)\textit{Duffy
  }}]{Duffy08}
 Duffy, T., (2008), Some Recent Advances in Understanding the Mineralogy of Earth’s Deep Mantle, \textit{Phil. Trans. Royal Soc. A}, \textit{366}, 4273--4293.

\bibitem[{\textit{Goncharov et~al.}(2006)\textit{Goncharov, Struzhkin, and Jacobse
  }}]{Goncharov} 
 Goncharov, A.~F., V.~V.~Struzhkin, and S.~D.~Jacobsen (2006), Reduced Radiative Conductivity of Low-Spin (Mg,Fe)O in the Lower Mantle, \textit{Science}, \textit{312}, 1205--1208.

 \bibitem[{\textit{Himmetoglu et~al.}(2014)\textit{Himmetoglu, Floris, de Gironcoli, and Cococcioni
  }}]{Himmetoglu14} 
 Himmetoglu, B., A.~Floris, S.~de Gironcoli, and M.~Cococcioni (2014), Hubbard-Corrected DFT Energy Functionals: The LDA+U Description of Correlated Systems, \textit{Int. J. Quantum Chem}, \textit{114}, 14--49.
 
 
\bibitem[{\textit{Hirose et~al.}(2005)\textit{Hirose, K., Kawamura, K., Ohishi, Y., Tateno, S.,  Sata, N.
  }}]{Hirose05} 
 Hirose, K.,  K. Kawamura, Y. Ohishi,  S. Tateno, N. Sata (2005), Stability and Equation of State of MgGeO$_3$ Post-Perovskite Phase, \textit{American Mineralogist}, \textit{90}, 262--265. 
 
 
 \bibitem[{\textit{Hsu et~al.}(2011)\textit{Han Hsu, Blaha, Cococcioni, and Wentzcovitch
  }}]{Hsu11} 
 Hsu, H., P.~Blaha, M.~Cococcioni, and R.~M.~Wentzcovitch (2011),  Spin-state Crossover and hyperfine interactions of ferric iron in MgSiO$_3$ Perovskite, 
 \textit{Phys. Rev. Lett.}, \textit{106}, 118501.
 
\bibitem[{\textit{Hsu et~al.}(2009)\textit{Han Hsu, Umemoto, Cococcioni, and Wentzcovitch
  }}]{Hsu09} 
 Hsu, H., K.~Umemoto, M.~Cococcioni, and R.~M.~Wentzcovitch (2009),  First principles study of low-spin LaCoO$_3$ with structurally consistent Hubbard U, 
 \textit{Phys. Rev. B}, \textit{79}, 125124.

\bibitem[{\textit{Hsu et~al.}(2010)\textit{Han Hsu, Umemoto, Wentzcovitch, and Blaha
  }}]{Hsu10} 
 Hsu, H., K.~Umemoto, R.~M.~Wentzcovitch, and P.~Blaha (2010),  Spin states and hyperfine interactions of iron in (Mg,Fe)SiO$_3$ perovskite under pressure, 
 \textit{Earth Planet. Sci. Lett.}, \textit{294}, 19--26.

 \bibitem[{\textit{Hsu et~al.}(2012)\textit{H. Hsu, Y. Yu, and R. M. Wentzcovitch
  }}]{Hsu12} 
 Hsu, H., Y. Yu, and R. M. Wentzcovitch (2012),  Effects of aluminum on spin-state crossover of iron in the Earth's lower mantle, 
 \textit{Earth Planet. Sci. Lett.}, \textit{359--360}, 34--39.
 
 \bibitem[{\textit{Hsu and Wentzcovitch}(2014)\textit{Han Hsu, and Wentzcovitch
  }}]{Hsu14} 
 Hsu, H., Wentzcovitch, R.~M., (2014), First-principles study of intermediate-spin ferrous iron in the Earth's lower mantle, 
 \textit{Phys. Rev. B}, \textit{90(19)}, 195205.
 
 
\bibitem[{\textit{Hohenberg and Kohn}(1964)\textit{Hohenberg, Kohn
  }}]{Hohenberg} 
 Hohenberg, P., and W.~Kohn (1964),  Inhomogeneous electron gas, \textit{Phys. Rev. B}, \textit{136}, 864--871.
 
 
\bibitem[{\textit{Ito et~al.}(2010)\textit{Ito, E., Yamazaki, D., Yoshino, T., Fukui, H., Zhai, S., Shatzkiy, A., Katsura, T., Tange, Y., and Funakoshi, K
  }}]{Ito10}
Ito, E., Yamazaki, D., Yoshino, T., Fukui, H., Zhai, S., Shatzkiy, A., Katsura, T., Tange, Y., Funakoshi, K., (2010), Pressure Generation and Investigation of the Post-Perovskite Transformation in MgGeO$_3$
by Squeezing the Kawai-Cell Equipped with Sintered Diamond Anvils, \textit{Earth Planet. Sci. Lett.}, \textit{293}, 84--89.
  
\bibitem[{\textit{Jackson et~al.}(2005)\textit{jackson, Sturhahn, Shen, Zhao, Hu, Errandonea,  Bass, and Fei
  }}]{Jackson05}
 Jackson, J.M., W.~Sturhahn, G.~Shen, J.~Zhao, M.~Y.~Hu, D.~Errandonea, J.~D.~ Bass, and Y.~Fei (2005), A synchrotron Mössbauer spectroscopy study of (Mg,Fe)SiO$_3$ perovskite up to 120 GPa,
 \textit{Am. Mineral.}, \textit{90},  199--205.

\bibitem[{\textit{Karki et~al.}(1999)\textit{Karki, B.~B. R. M. Wentzcovitch. S. de Gironcoli, and S. Baroni
  }}]{Karki99} 
Karki, B.~B., R. M. Wentzcovitch. S. de Gironcoli, and S. Baroni (1999),  First principles determination elastic anisotropy and wave velocities of MgO at lower mantle conditions, \textit{Science}, \textit{286}, 1705.

\bibitem[{\textit{Karkiet~al.}(2000)\textit{Karki, B.~B. R. M. Wentzcovitch. S. de Gironcoli, and S. Baroni
  }}]{Karki00} 
Karki, B.~B., R. M. Wentzcovitch. S. de Gironcoli, and S. Baroni (2000),   \textit{Ab initio} lattice dynamics of MgSiO$_3$-perovskite, \textit{Phys. Rev. B}, \textit{62}, 14750.

\bibitem[{\textit{Kohn and Sham}(1964)\textit{Kohn and Sham
  }}]{Kohn} 
 Kohn, W., and L.~J.~Sham (1964),  Self‐Consistent equations including
exchange and correlation effects, \textit{Phys. Rev. A}, \textit{140}, 1133--1138.

\bibitem[{\textit{Kubo et~al.}(2006)\textit{Kubo, A., Kiefer, B., Shen, G., Prakapenka, V., Cava, R., and Duffy, T.
  }}]{Kubo06}
Kubo, A., B.~Kiefer, G. Shen, V. Prakapenka, R. Cava, and T. Duffy (2006), Stability and Equation of
State of the Post-Perovskite Phase in MgGeO$_3$ to 2 Mbar, \textit{Geophys. Res. Lett.}, \textit{33}, L12812.


\bibitem[{\textit{Kulik et~al.}(2006)\textit{Kulik, Cococcioni, Scherlis, and Marzari
  }}]{Kulik}
Kulik, H., M.~Cococcioni, D.~A.~Scherlis, and N.~Marzari (2006), Density functional theory
in transition metal chemistry: a self-consistent Hubbard U approach, \textit{Phys. Rev. Lett.}, \textit{97}, 103001.

\bibitem[{\textit{Leinenweber et~al.}(2004)\textit{Leinenweber, K., Wang, Y. B., Yagi, T., and Yusa, H.
  }}]{Leinenweber94}
Leinenweber, K., Y. B. Wang,  T. Yagi, and H. Yusa (1994), An Unquenchable Perovskite Phase of MgGeO$_3$ and Comparison with MgSiO$_3$ Perovskite,
\textit{American Mineralogist}, \textit{179}, 197-199.


\bibitem[{\textit{Li et~al.}(2004)\textit{Li,  Struzhkin, Mao, Shu,  Hemley, Fei, Mysen, Dera, Prakapenka, and Shen
  }}]{Li04}
Li, J., V.~V.~Struzhkin, H.~K.~Mao, J.~Shu, R.~J.~Hemley, Y.~Fei, B.~Mysen, P.~Dera, V.~Prakapenka, and G.~Shen (2004), Electronic spin state of iron in lower mantle perovskite,
\textit{Proc. Natl. Acad. Sci.}, \textit{101}, 14027--14030.

\bibitem[{\textit{Li et~al.}(2006)\textit{Li, Sturhahn, Jackson, Struzhkin, Lin, Zhao, Mao, and Shen
  }}]{Li06}
Li, J., W.~Sturhahn, J.~M.~Jackson, V.~V.~Struzhkin, J.~F.~Lin, J.~Zhao, H.~K.~Mao, and G.~Shen (2006), Pressure effect on the electronic structure of iron in (Mg,Fe)(Si,Al)O$_3$
perovskite: A combined synchrotron Mössbauer and X-ray emission spectroscopy study up to 100 GPa, 
\textit{Phys. Chem. Minerals}, \textit{33}, 575--585.

\bibitem[{\textit{Lin et~al.}(2008)\textit{Lin, Watson, Vanko, Alp, Prakapenka, Dera, Struzhkin, Kubo, Zhao,  McCammon, and Evans
  }}]{Lin08}
 Lin, J., H.~Watson, G.~Vanko, E.~E.~Alp, V.~B.~Prakapenka, P.~Dera, V.~V.~Struzhkin, A.~Kubo, J.~Zhao, C.~McCammon, and W.~J.~Evans (2008), Intermediate-spin ferrous iron in
lowermost mantle post-perovskite and perovskite, \textit{Natl. Geosci}, \textit{1}, 688--691.

\bibitem[{\textit{Lin et~al.}(2012)\textit{Lin, Alp, Mao, Inoue, McCammon, Xiao, Chow, and Zhao
  }}]{Lin12}
 Lin, J.~F., E.~E.~Alp, Z.~Mao, T.~Inoue, C.~McCammon, Y.~Xiao, P.~Chow, and J.~Zhao (2012), Electronic spin and valence states of iron in the lower-mantle
silicate perovskite by synchrotron Mössbauer spectroscopy, \textit{American Mineralogist}, \textit{97}, 592--597.


\bibitem[{\textit{Lin et~al.}(2013)\textit{J. F. Lin, S. Speziale, Z. Mao, and H. Marquardt
  }}]{Lin13}
Lin, J.~F., S. Speziale,  Z. Mao, and H. Marquardt (2013), Effects of the electronic spin transitions of iron in lower-mantle minerals: implications to deep-mantle geophysics and geochemistry,
\textit{ Rev. Geophys.}, \textit{51}, 244--275.


\bibitem[{\textit{Mao et~al.}(2004)\textit{Mao, W.L., Shen, G., Prakapenka, V.B., Meng, Y., Campbell, A.J., Heinz, D.L., Shu, J., Hemley, R.J., Mao, H.-k.,
  }}]{Mao04}
 Mao, W.L., G.~Shen, V.~B.~Prakapenka, Y. Meng, A.~J.~Campbell, D.~L.~Heinz, J.~Shu, R.~J.~Hemley,and  H.-K. Mao (2004), Ferromagnesian postperovskite silicates in the D'' layer of the earth. \textit{Proc. Natl. Acad. Sci.}, \textit{101(45)}, 15867--15869.


\bibitem[{\textit{Mao et~al.}(2010)\textit{Mao, Z., Lin, J.F., Jacobs, C., Watson, H.C., Xiao, Y., Chow, P., Alp, E.E., Prakapenka, V.B.
  }}]{Mao10}
 Mao, Z., J.F.~Lin, C. Jacobs, H.~C. Watson, Y. Xiao, P. Chow, E.~E. Alp, and V.~B. Prakapenka (2008), Electronic spin and valence states of Fe in CaIrO$_3$-type silicate post-perovskite
 in the Earth's lowermost mantle, \textit{Geophys. Res. Lett.}, \textit{37}, L22304.


\bibitem[{\textit{Marquardt et~al.}(2009)\textit{Marquardt, Speziale, Reichmann, Frost, Schilling, Garnero
  }}]{Marquardt09a}
Marquardt M., S. Speziale, H.~J. Reichmann, D.~J. Frost, F.~R. Schilling, and E.~J. Garnero (2009),  Elastic Shear Anisotropy of Ferropericlase in Earth's Lower Mantle, \textit{Science}, \textit{324}, 224--226.

\bibitem[{\textit{McCammon et~al.}(2008)\textit{McCammon, Kantor, Narygina, Rouquette, Ponkratz, Sergueev, Mezouar, Prakapenka, and Dubrovinsky
  }}]{McCammon08}
McCammon, C., I.~Kantor, O.~Narygina, J.~Rouquette, U.~Ponkratz, I.~Sergueev, M.~Mezouar, V.~Prakapenka, and L.~Dubrovinsky (2008), Stable intermediate-spin
ferrous iron in lower-mantle perovskite, \textit{Nat. Geosci.}, \textit{1}, 684--687.

\bibitem[{\textit{McCammon et~al.}(2010)\textit{McCammon, C., Dubrovinsky, L., Narygina, O., Kantor, I., Wu, X., Glazyrin, K., Sergueev, I., Chumakov, A.I.
  }}]{McCammon10}
McCammon, C., L.~Dubrovinsky, O.~Narygina, I.~Kantor,  X.~Wu, K.~Glazyrin, I.~Sergueev, and A.~I Chumakov (2008), Low-spin Fe$^{2+}$ in silicate perovskite and a possible layer at
the base of the lower mantle, \textit{Phys. Earth Planet. Inter.}, \textit{180}, 215--221.


\bibitem[{\textit{Merkel et~al.}(2006)\textit{Merkel, S., Kubo, A., Miyagi, L., Speziale, S., Duffy, T., Mao, H., Wenk, H.
  }}]{Merkel06}
Merkel, S., A.~Kubo, L. Miyagi, S. Speziale, T. Duffy, H. Mao, and H. Wenk (2006), Deformation of MgGeO$_3$ post-Perovskite at Lower Mantle Pressures, \textit{Science}, \textit{311}, 644--646.

\bibitem[{\textit{Monkhorst and Pack}(1976)\textit{Kulik, Cococcioni, Scherlis, and Marzari
  }}]{Monkhorst_Pack}
Monkhorst, H.~J., and J.~D. Pack (1976),  Special points for Brillouin-zone integrations, \textit{Phys. Rev. B}, \textit{13}, 5188--5192.


\bibitem[{\textit{Murakami et~al.}(2004)\textit{Murakami, Hirao, kawamura and Hirose
  }}]{Murakami04}
Murakami, M., K.~Hirose, K.~Kawamura, N.~Sata, Y.~Ohishi (2004), Post-perovskite phase transition in MgSiO$_3$ data, \textit{Science}, \textit{304}, 855--858.

\bibitem[{\textit{Nunez-Valdez et~al.}(2012a)\textit{Nunez-Valdez, Wu, Yu, Revenaugh, and Wentzcovitch
  }}]{Nunez-Valdez12a}
Nunez-Valdez, M., Z. Wu, Y.~G.~Yu, J.~Revenaugh, and R.~M.~Wentzcovitch (2012a), Thermoelastic properties of ringwoodite (Fe$_x$Mg$_{1-x}$)$_2$SiO$_4$: Its relationship
to the 520 km seismic discontinuity , \textit{Earth Planet.
Sci. Lett.}, \textit{351-352}, 115--122.

\bibitem[{\textit{Nunez-Valdez et~al.}(2012b)\textit{Nunez-Valdez, Umemoto, and Wentzcovitch
  }}]{Nunez-Valdez12b}
Nunez-Valdez, M., K.~Umemoto, and R.~M.~Wentzcovitch (2012b), Elasticity of diamond at high pressures and temperatures  , \textit{Appl. Phys. Lett.}, \textit{101}, 170912.

\bibitem[{\textit{Nunez-Valdez et~al.}(2013)\textit{Nunez-Valdez, Wu, Yu, and Wentzcovitch
  }}]{Nunez-Valdez13}
Nunez-Valdez, M., Z. Wu, Y.~G.~Yu, and R.~M.~Wentzcovitch (2013), Thermal elasticity of (Fe$_x$Mg$_{1-x}$)$_2$SiO$_4$ olivine and wadsleyite, \textit{Geophys. Res. Lett.}, \textit{40}, 290--294.

\bibitem[{\textit{Oganov and Ono}(2004)\textit{Oganov, A.R., Ono, S.
  }}]{Oganov04}
Oganov, A.R., and S. Ono (2004), Theoretical and experimental evidence for a post-perovskite phase of MgSiO$_3$ in Earth's D`` layer, \textit{Nature}, \textit{430}, 445--448.


\bibitem[{\textit{Perdew et~al.}(1996)\textit{Perdew, J.~P.,  Burke, K., Ernzerhof, M.,
  }}]{Perdew96} 
 Perdew, J.~P.,  K.~Burke, and M. Ernzerhof (1996),  Generalized Gradient Approximation Made Simple, \textit{Phys. Rev. Lett.}, \textit{77}, 3865.
 
\bibitem[{\textit{Ross and Navrotsky}(1988)\textit{Ross, N.~L.,  Navrotsky, A.,
  }}]{Ross88} 
Ross, N.~L., and A. Navrotsky (1988), Study of the MgGeO$_3$ Polymorphs (ortho-Pyroxene, Clinopyroxene, and Ilmenite Structures) by Calorimetry, Spectroscopy, and Phase-Equilibria,
\textit{American Mineralogist}, \textit{73}, 1355--1365.

\bibitem[{\textit{Runge et~al.}(2006)\textit{Runge, C., Kubo, A., Kiefer, B., Meng, Y., Prakapenka, V., Shen, G., Cava, R., and Duffy, T.,
  }}]{Runge06} 
Runge, C., A. Kubo, B. Kiefer, Y. Meng, V. Prakapenka, G. Shen, R. Cava, and T. Duffy (2006), Equation of State of MgGeO$_3$ Perovskite to 65 GPa: Comparison with the Post-Perovskite Phase,
\textit{Phys. and Chem. Minerals}, \textit{33}, 699--709.

\bibitem[{\textit{Shieh et~al.}(2006)\textit{Shieh, S.R., Duffy, T.S., Kubo, A., Shen, G., Prakapenka, V.B., Sata, N., Hirose, K., Ohishi, Y.,
  }}]{Shieh06} 
Shieh, S.R., T.~S. Duffy, A. Kubo, G. Shen, V.~B. Prakapenka, N. Sata, K. Hirose, and Y. Ohishi (2006), Equation of state of the postperovskite phase synthesized from a natural (Mg,Fe)SiO$_3$ orthopyroxene,
\textit{Proc. Natl. Acad. Sci.}, \textit{103(9)}, 3039-3043.


\bibitem[{\textit{Shim et~al.}(2007)\textit{Shim, S., Kubo, A., and Duffy, T.,
  }}]{Shim07} 
Shim, S., A. Kubo, and T. Duffy (2007), Raman Spectroscopy of Perovskite and Post-Perovskite Phases of MgGeO$_3$ to 123 GPa,
\textit{Earth Planet. Sci. Lett.}, \textit{260}, 166--178.


\bibitem[{\textit{Stackhouse et~al.}(2006)\textit{Stackhouse, S., Brodholt, J.P., Price, G.D.,
  }}]{Stackhouse06} 
Stackhouse, S., J.~P.~Brodholt, and G.~D. Price (2006),  Elastic anisotropy of FeSiO$_3$ end-members of the perovskite and post-perovskite phases,
\textit{Geophys. Res. Lett.}, \textit{33}, L01304.

\bibitem[{\textit{Tsuchiya et~al.}(2004)\textit{Tsuchiya, T., Tsuchiya, J., Umemoto, K., Wentzcovitch, R.M.
  }}]{Tsuchiya04} 
Tsuchiya, T., J.~Tsuchiya, K.~Umemoto, and R.~M. Wentzcovitch, (2004),   Phase transition in MgSiO$_3$ perovskite in the earth's lower mantle, \textit{Earth Planet. Sci. Lett.}, \textit{224}, 241-248.
 
\bibitem[{\textit{Tsuchiya et~al.}(2006)\textit{Tsuchiya, T., Wentzcovitch, R.M., da Silva, C.R.S., de Gironcoli, S.,
  }}]{Tsuchiya06} 
Tsuchiya, T., R.~M. Wentzcovitch, C.~R.~S.~da Silva, and S.~de Gironcoli (2006),   Spin transition in magnesiowüstite in Earth's lower mantle, \textit{Phys. Rev. Lett.}, \textit{96(19)}, 198501.
 
\bibitem[{\textit{Tsuchiya and Tsuchiya}(2007)\textit{Tsuchiya, T., Tsuchiya, J.,
  }}]{Tsuchiya07} 
Tsuchiya, T., and J.~Tsuchiya (2007),  High-Pressure-High-Temperature Phase Relations of MgGeO$_3$: First-Principles Calculations, \textit{Phys. Rev. B}, \textit{76}, 092105.


\bibitem[{\textit{Vanderbilt}(1990)\textit{Vanderbilt
  }}]{Vanderbilt}
Vanderbilt, D., (1990), Soft self-consitent pseudopotentials in a genrelized eigenvalue formalism, \textit{Phys. Rev. B}, \textit{41}, 7892--7895.


\bibitem[{\textit{Wentzcovitch}(1991)\textit{Wentzcovitch
  }}]{Wentzcovitch91}
Wentzcovitch, R.~M., (1991), Invariant molecular dynamics approach to structural phase transitions, \textit{Phys. Rev. B}, \textit{44}, 2358-2361.


\bibitem[{\textit{Wentzcovitch et~al.}(1993)\textit{Wentzcovitch, Martins, and Price
  }}]{Wentzcovitch93}
Wentzcovitch, R.~M., J.~L.~Martins, and G.~D.~Price (1993), \textit{Ab initio} molecular dynamics with variable cell shape: application to MgSiO$_3$, \textit{Phys. Rev. Lett.}, \textit{70}, 3947.

\bibitem[{\textit{Wentzcovitch et~al.}(2004)\textit{Wentzcovitch, Karki, Cococcioni, and de Gironcoli
  }}]{Wentzcovitch04}
Wentzcovitch, R.~M., B.~B. Karki, M. Cococcioni, and S. de Gironcoli (2004),  Thermoelastic properties of of
MgSiO$_3$-perovskite: insights on the nature of the Earth’s lower mantle, \textit{Phys. Rev. Lett.}, \textit{92}, 018501.

\bibitem[{\textit{Wentzcovitch et~al.}(2009)\textit{Wentzcovitch, Justo, Wu, da Silva, Yuen, and Kohlstedtd
  }}]{Wentzcovitch09}
Wentzcovitch, R.~M., J.~F.~Justo, Z.~Wu, C.~R.~S~da Silva, A.~Yuen, and D.~Kohlstedtd (2009), Anomalous compressibility of ferropericlase throughout the iron spin cross-over, \textit{Proc. Natl. Acad. Sci.}, \textit{106}, 21.


\bibitem[{\textit{Wu et~al.}(2009)\textit{Wu, Justo, da Silva, de Gironcoli, and Wentzcovitch
  }}]{Wu09}
Wu, Z., J.~F.~Justo, C.~R.~S.~da Silva, S.~de Gironcoli, and R.~M.~Wentzcovitch (2009), Anomalous thermodynamic properties in ferropericlase throughout its spin crossover, \textit{Phys. Rev. B}, \textit{80}, 014409.


\bibitem[{\textit{Wu et~al.}(2013)\textit{Wu, Justo and Wentzcovitch
  }}]{Wu13}
Wu, Z., J.~F.~Justo, and R.~M.~Wentzcovitch (2013), Elastic Anomalies in a Spin-Crossover System: Ferropericlase at Lower Mantle Conditions, \textit{Phys. Rev. Lett.}, \textit{110}, 228501.


\bibitem[{\textit{Wu and Wentzcovitch}(2014)\textit{Wu and Wentzcovitch
  }}]{Wu14}
Wu, Z., and R.~M.~Wentzcovitch (2014), Spin crossover in ferropericlase and velocity heterogeneities in the lower mantle, \textit{Proc. Natl. Acad. Sci.},  \textit{111}, 29

\bibitem[{\textit{Yu et~al.}(2012)\textit{Yu, Y., Hsu,  H., Cococcioni, M.,  and  Wentzcovitch, R.~M.,
  }}]{Yu12}
Yu, Y.~G., H.~Hsu,  M. Cococcioni, and R.~M. Wentzcovitch (2012), Spin states and hyperfine interactions of iron incorporated in MgSiO$_3$ post-perovskite, \textit{Earth Planet. Sci. Lett.},
\textit{331-332}, 1-7.
 

\bibitem[{\textit{Zhang and Oganov}(2006)\textit{Zhang, F., Oganov, A.R
  }}]{Zhang_and_Oganov06}
Zhang, F., and A.~R. Oganov (2006),  Valence state and spin transitions of iron in Earth's mantle silicates, \textit{Earth Planet. Sci. Lett.},  \textit{249}, 436-443.
 
\end{thebibliography}
\end{document}